\documentclass[global,final]{svjour}

\usepackage{graphicx,rotating}

\newif\ifPDF
    \ifx\pdfoutput\undefined
      \PDFfalse
    \else
      \PDFtrue
    \fi
    \ifPDF
      \usepackage[pdftex]{hyperref}
    \else
      \usepackage[dvips]{graphicx}
      \newcommand{\href}[2]{#2}
    \fi

\begin{document}

\title{Evaluating the performance of adapting trading strategies with different memory lengths}
\author{Andreas Krause}

\institute{University of Bath, School of Management,Bath BA2 7AY, Great Britain, \email{mnsak@bath.ac.uk}}

\date{Received: / Accepted: }

\maketitle

\begin{abstract}
  We propose a prediction model based on the minority game in which traders continuously evaluate a complete set of trading strategies with different memory lengths using the strategies' past performance. Based on the chosen trading strategy they determine their prediction of the movement for the following time period of a single asset.  We find empirically using stocks from the S\&P500 that our prediction model yields a high success rate of over 51.5\% and produces higher returns than a buy-and-hold strategy. Even when taking into account trading costs we find that using the predictions will generate superior investment portfolios.

\end{abstract}

\section{Introduction}

Traders in financial markets continuously adapt their trading behavior to the changing market conditions and different composition of traders active in the market. Given these changes, it is reasonable to expect that any trading rule which has proved to be successful in the past will not necessarily be so in the future; thus traders will have to switch between trading strategies. A large amount of literature on such trading strategies, usual based on technical analysis, finds very mixed evidence on the profitability of such strategies with results very much depending on the assets chosen and time periods investigated, see \cite{Fernandezetal00}, \cite{Allen/Karjalainen99}, \cite{Marshalletal06}, \cite{Nametal05}, \cite{Brocketal92}, and \cite{Loetal00}, among many others. These mixed results suggest that some strategies work at times but not at others and ideally traders should change the trading strategy they are choosing, however thus far no such endogenous switch between trading strategies has been investigated in detail. \cite{Chenetal08} allow for changing trading strategies, also using the minority game as a basis, but they employ a different mechanism for their predictions than in our contribution.

In the past the minority game has been shown to replicate the properties of asset markets particularly well, see \cite{Challet/Zhang97}, \cite{Challet/Zhang98}, \cite{Challetetal00}, \cite{Challetetal01} for an overview. In the minority game an odd number of traders repeatedly choose between buying and selling an asset. In order to be profitable, the traders seek to be in the minority, i.e. buy when the majority of traders are selling and sell when the majority of traders are buying. In order to achieve this goal, traders seek to predict whether in the next time period the majority of traders will be buying or selling. They conduct this prediction using the aggregate outcomes from the past $M$ time periods (the history) and depending on the pattern observed make their choice whether to buy or sell. There exist $2^{2^M}$ different ways to conduct these predictions, called "strategies", and the traders continuously keep score of their performance in the past and will follow the strategy which shows the best past performance. Thus traders will change the strategy they are following, reacting to the performance of the strategy they are are using as well as the performance of all the alternative strategies. Hence traders will adapt their strategies to the changing trading environment and choose endogenously the best trading strategy.

We can use the set-up of the minority game to develop a framework for predicting movements of the stock market; if more traders are seeking to buy than sell, the price of the asset increases and falls otherwise. We can thus use the direction of past price movements as predictors for the direction of future price movements in the same way the minority game does. We would therefore employ a prediction method which endogenously changes the reaction to observed past price formation; this is in contrast to the technical analysis which does not adjust its treatment of price patterns to such observations.

In the coming section we will describe in more detail the forecasting mechanism used in this paper and in section 3 assess its qualities empirically. Section 4 then concludes the findings.

\section{The prediction mechanism}

We attempt to predict the sign of movements in asset markets one time period ahead. Employing the mechanism used for the minority game we collect information on the sign of the past $M$ price changes and use this as the basis for our prediction. Unlike the minority game, however, we do not set the memory length $M$ exogenously, but traders choose the optimal memory length endogenously. In order to make this choice, the traders keep record not only of the performance of each strategy for a given memory length but also compare the performance of the best strategy for each memory length. The trader chooses the best strategy of the memory length that provides the best performance at the time of the decision-making. Thus traders will change the memory length they are using, depending on the performance of the memory at the time.

We investigated the performance of this prediction mechanism using the 375 stocks from the S\&P500 that were continuously included in the index from 29 May 1991 to 29 May 2006 (3914 trading days) and we used all memory lengths up to $M=10$. As with increasing memory length the number of possible strategies increases substantially, e.g. for $M=10$ there are 1,048,576 possible strategies, we limit the number of strategies considered to 10,000; if there are more possible strategies, we select 10,000 strategies randomly.\footnote{We have investigated the stability of our results if we change the selection of strategies and found no meaningful differences in the outcomes.}

\section{Evaluation of the predictions}

\begin{figure}[h]
    \centering
    \rotatebox{0}{\includegraphics[width=120mm]{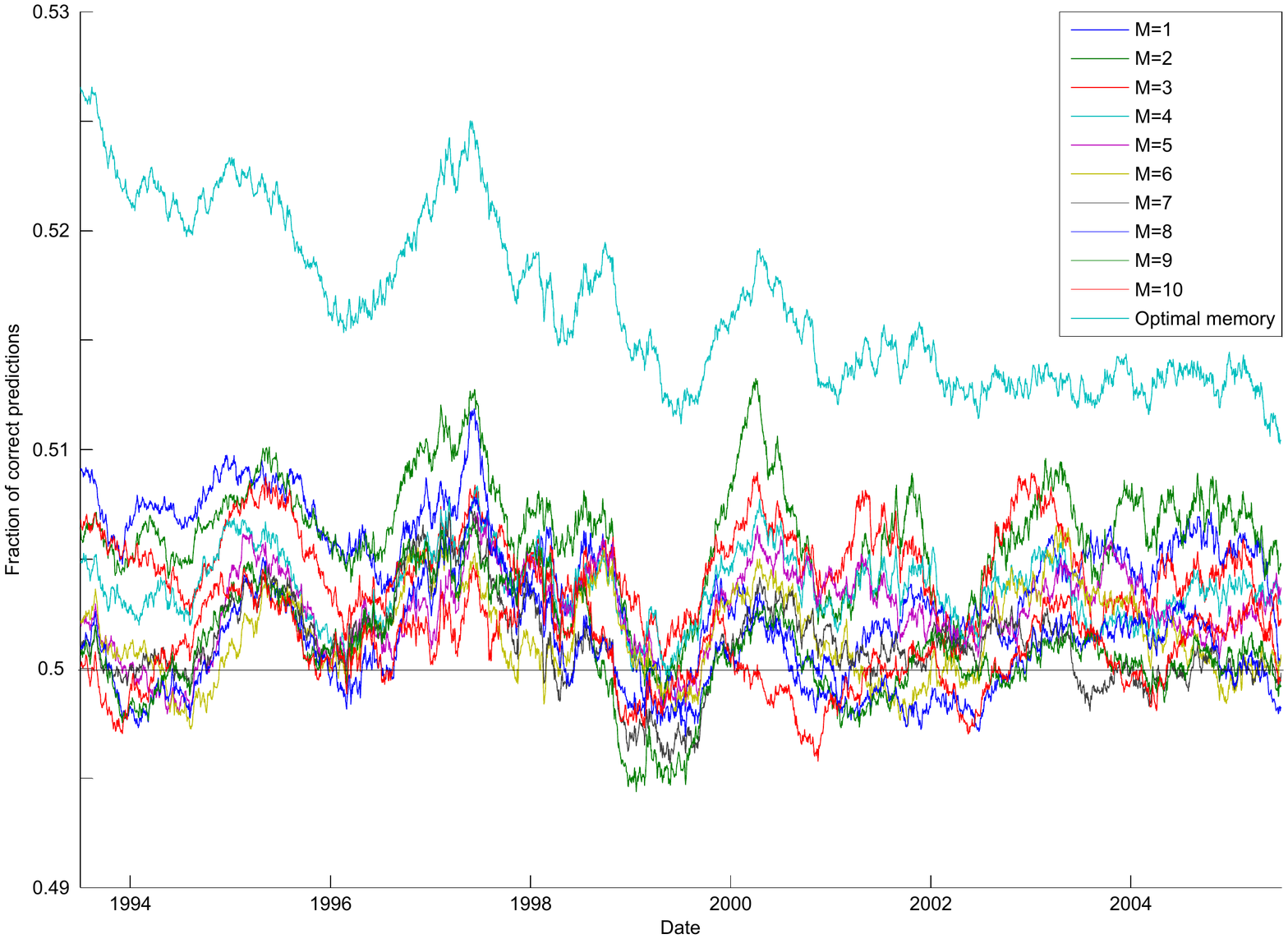}}
    \caption{Success rate predictions using a 250-trading day rolling window averages across all stocks for different memory lengths and the optimized memory length.}\label{fig1}
\end{figure}

We see quite clearly from figure \ref{fig1} that the success rate of our prediction exceeds 50\% significantly; this apparent success can be either the result of the predictions identifying some repeated patterns or alternatively that a trend in the prices is followed, e.g. an upward trend would give the prediction "increase" a success rate above 50\%. We find a small bias in favor of positive returns, they make 50.2\% of all observations and if we knew the movement of all stocks over the sample period in advance, i.e. whether they move up or down, the prediction would be correct 50.65\% of the time; both numbers are significantly smaller than the success rate of our predictions.

\begin{figure}
    \centering
    \rotatebox{0}{\includegraphics[width=120mm]{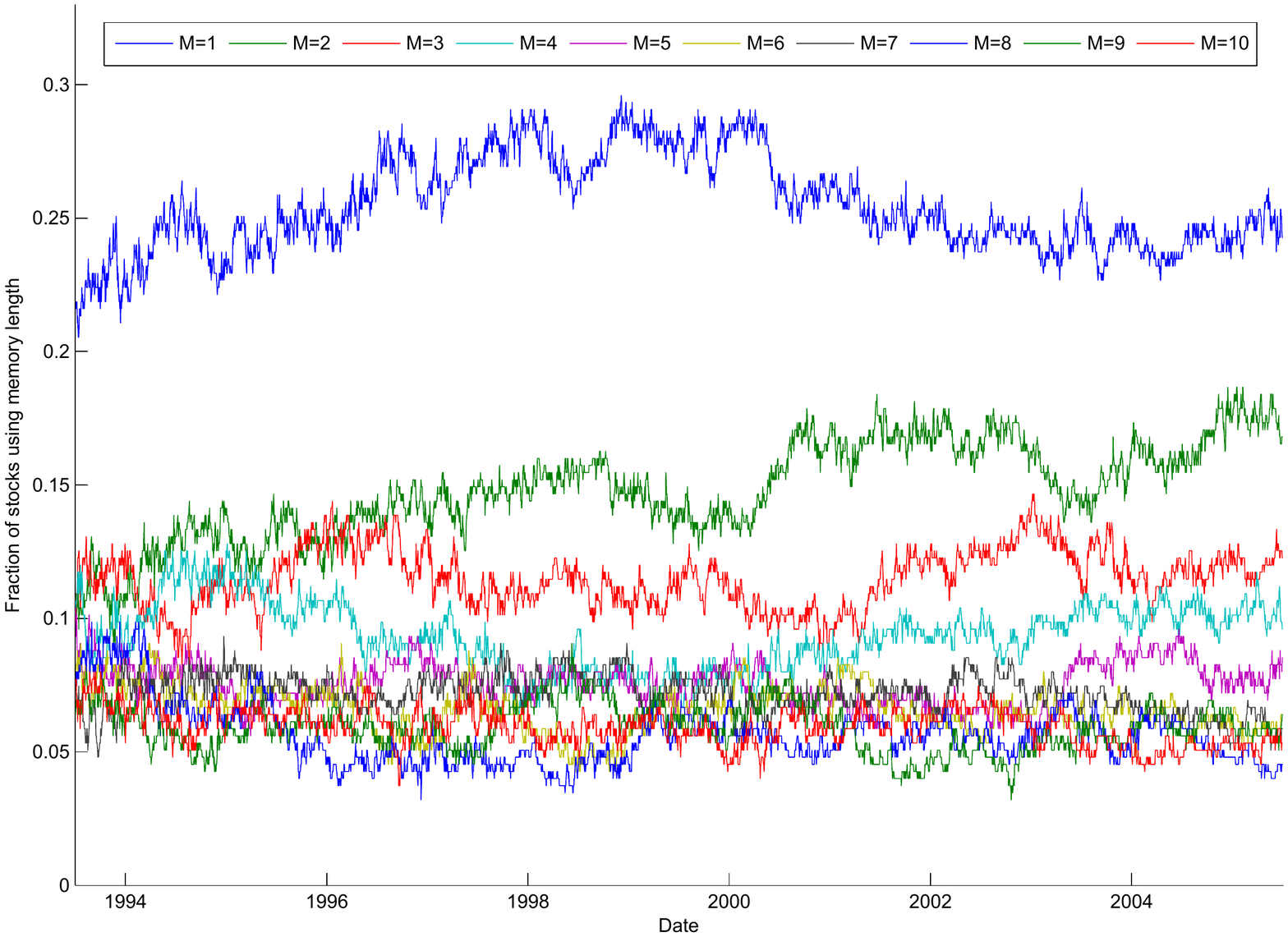}}
    \caption{Distribution of the optimal memory length.}\label{fig5}
\end{figure}

Neglecting the first 500 trading days in which the system learns the best predictions, we observe a consistently strong performance over 13 years with a success rate of approximately 51.5-52.0\% in any given year. We also see from figure 1 that allowing traders to switch the memory length they are using, significantly improves the performance compared to any of the fixed memory lengths. Looking at figure \ref{fig5} we observe that all memory lengths are actually used, although short memory lengths clearly dominate. This provides evidence for the existence of very short-term patterns in the data that our predictions exploit; longer memory lengths are not beneficial in general as the model attempts to detect patterns for the entire history length which do not exist.

\begin{figure}
    \centering
    \rotatebox{0}{\includegraphics[width=120mm]{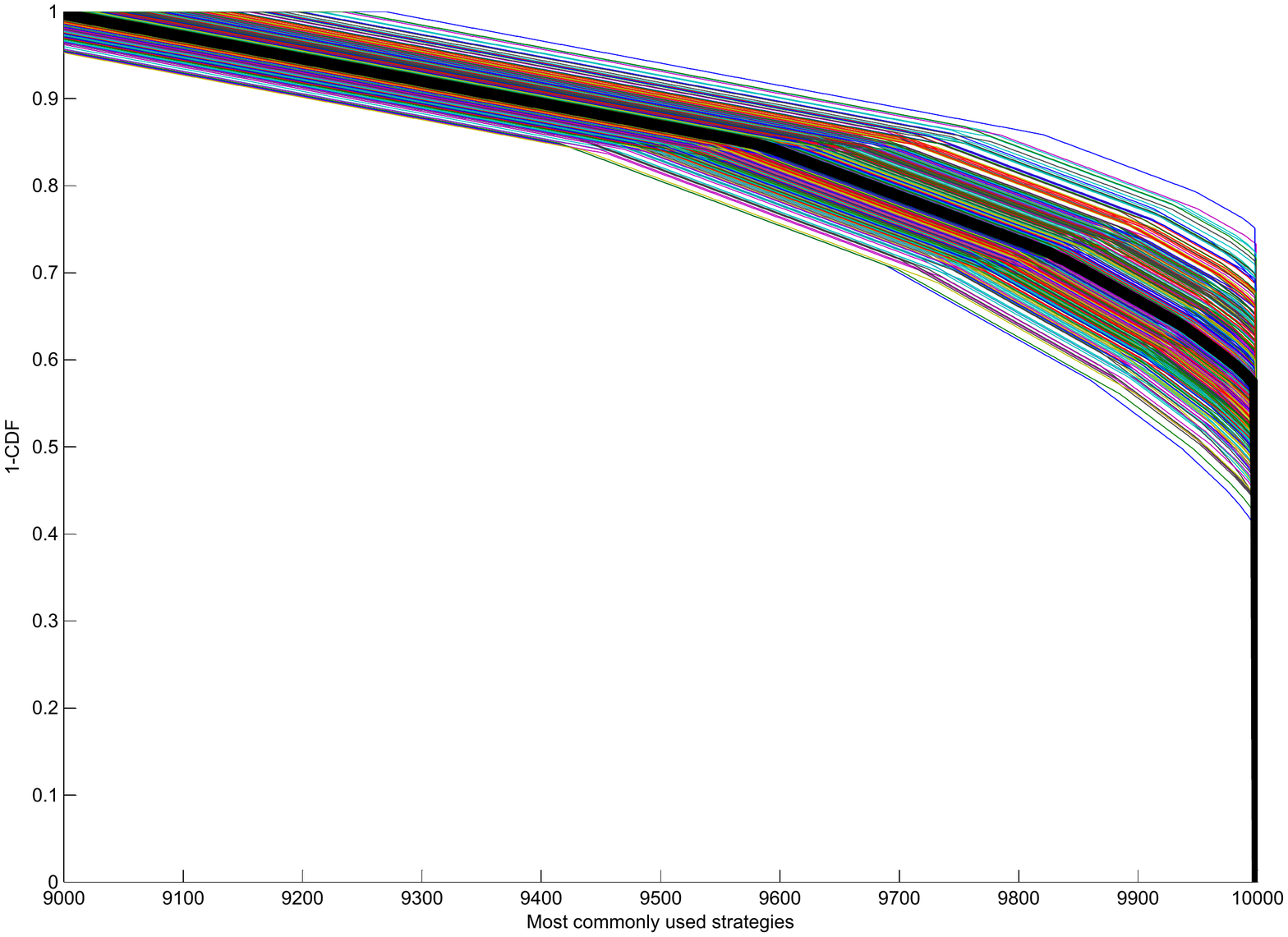}}
    \caption{Distribution function of the use of the number of optimal trading strategies for $M=10$ for all stocks with the median use in bold}\label{fig6}
\end{figure}

\begin{figure}
    \centering
    \rotatebox{0}{\includegraphics[width=120mm]{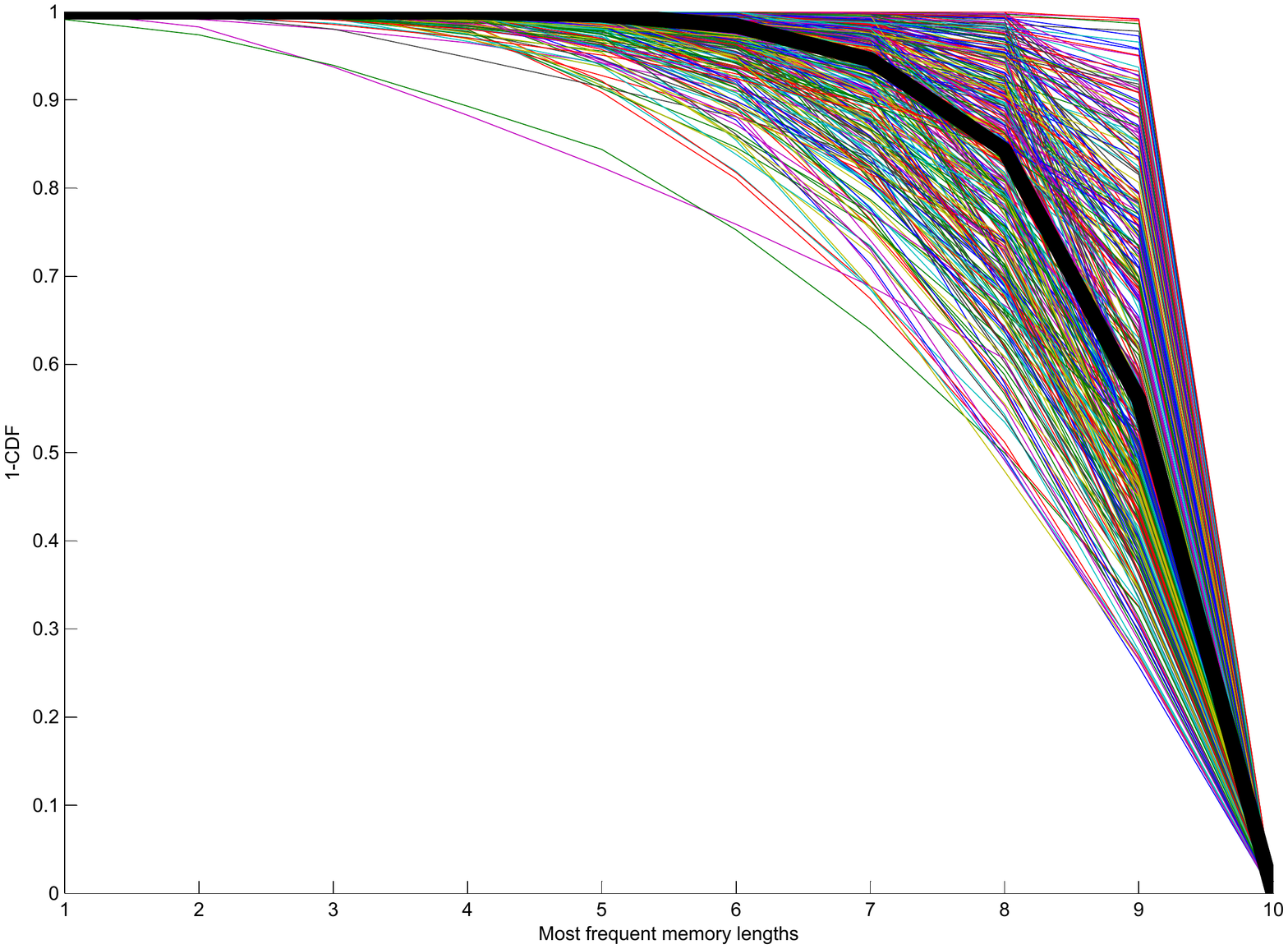}}
    \caption{Distribution function of optimal memory lengths used for all stocks with median use in bold.}\label{fig7}
\end{figure}

Furthermore we find that only a small number of trading strategies are actually chosen by the model, providing evidence for a relative stability of any patterns detected and successfully exploited. Figure \ref{fig6} shows the distribution of the strategies chosen in the case of $M=10$. We see very clearly that only a very small fraction of the available strategies are used by each individual stock, the most commonly chosen strategy being used about 50\% of the time, the second most common about 10\% of the time; in most cases less than 1,000 of the 10,000 available strategies get chosen at any point of time. Similarly we see from figure \ref{fig7} that only a small fraction of the memory lengths is used for each stock; about half the time is spent using the same memory and the three most frequently used memory lengths are employed about 90\% of the time.

\begin{figure}
    \centering
    \rotatebox{0}{\includegraphics[width=120mm]{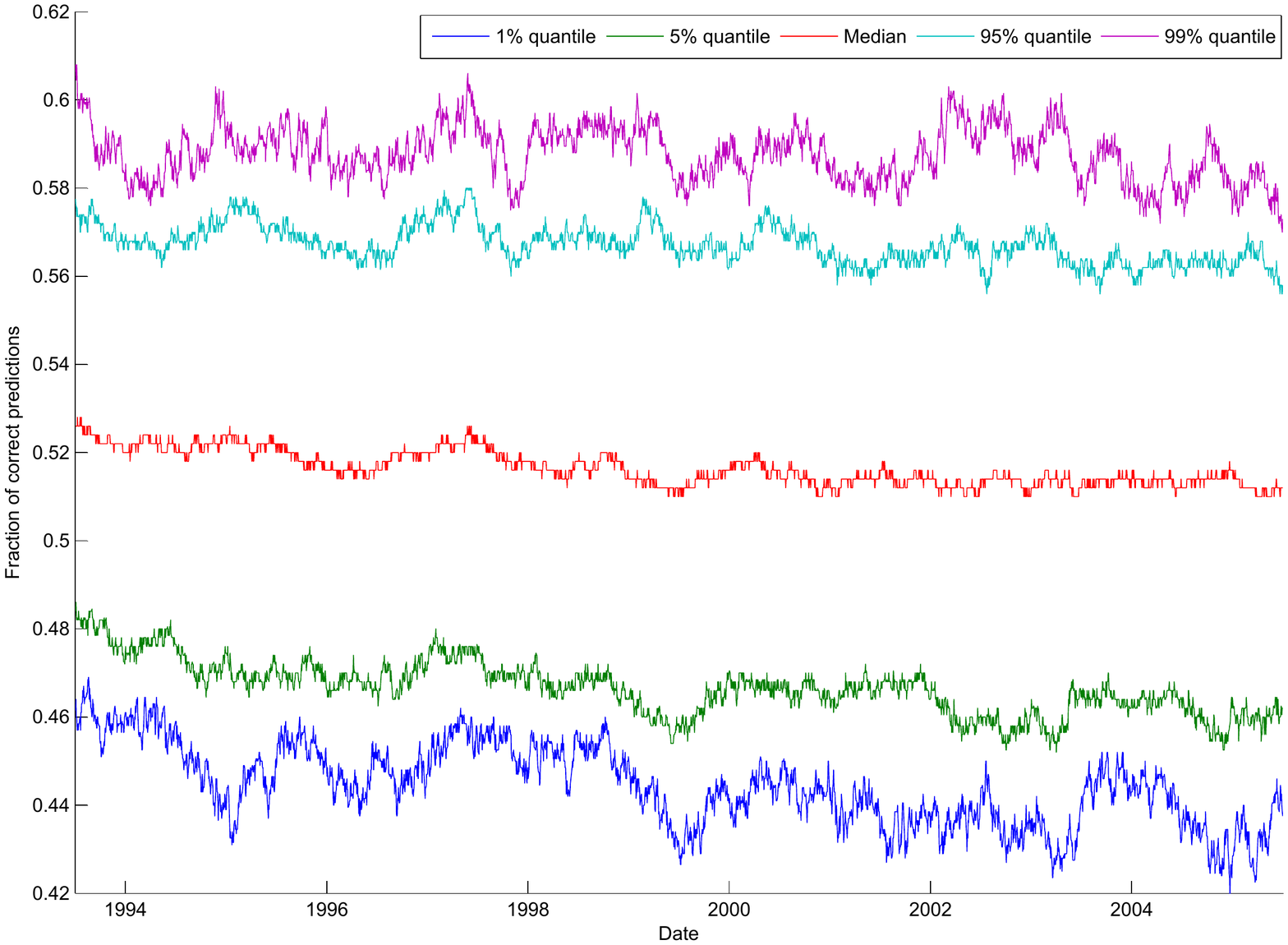}}
    \caption{Quantiles of the success rate of predictions using a 250-trading day rolling window for all stocks.}\label{fig2}
\end{figure}

It is, however, clear from figure \ref{fig2} that the success rate of predictions is below 50\% for a sizeable fraction of shares and performs significantly better for other stocks. With this result, it might be possible that following the predictions as proposed here might result in an overall lower performance than a simple buy-and-hold strategy.

\begin{figure}
    \centering
    \rotatebox{0}{\includegraphics[width=120mm]{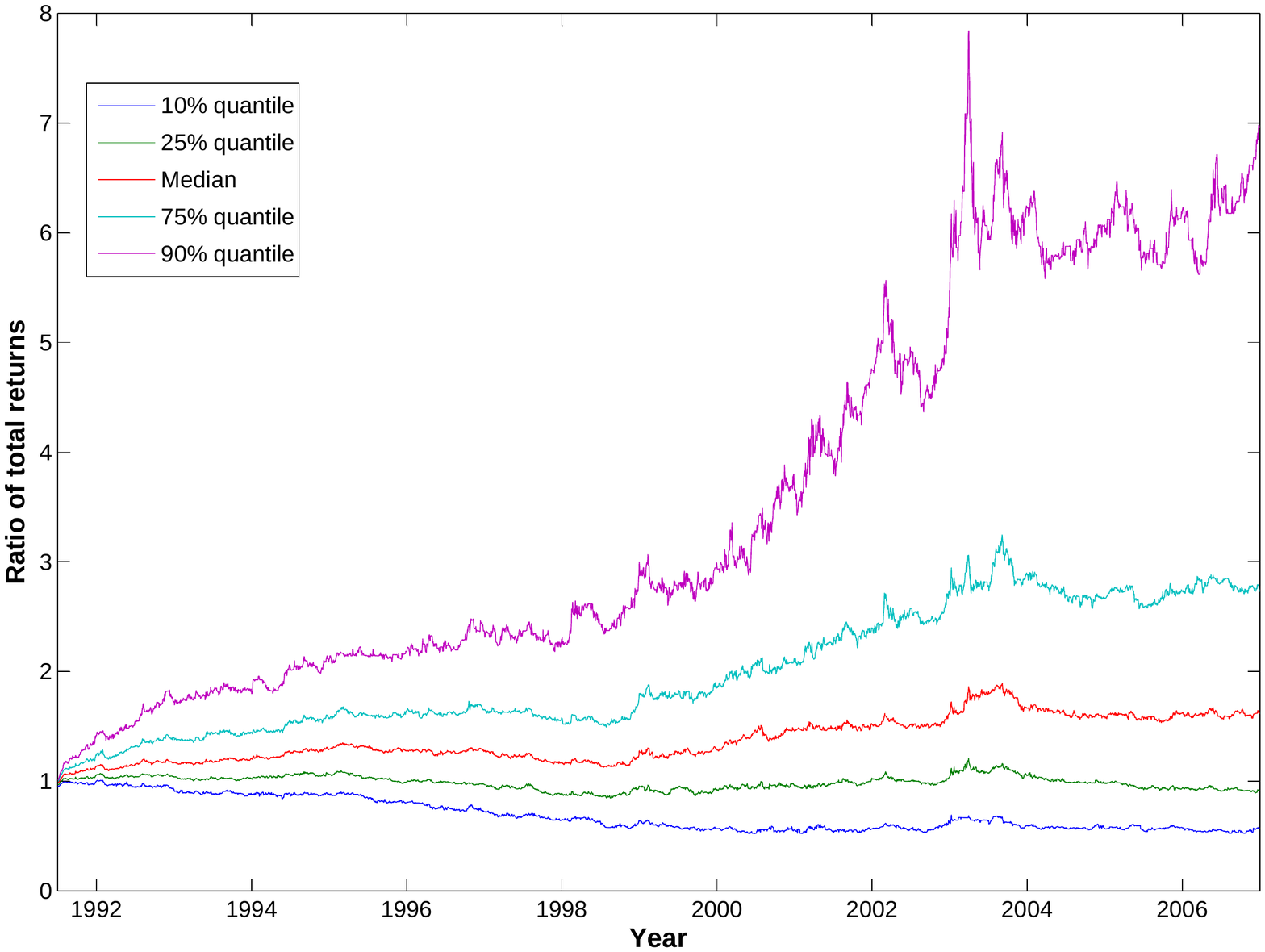}}
    \caption{Ratio of the total returns generated from our trading strategy, neglecting trading costs, and a buy-and-hold strategy from the start of the sample period. The figure shows the 10\%, 25\%, 75\%, and 90\% quantiles as well as the median of this ratio for the sample of 375 stocks considered. A value above 1 implies that our strategy outperforms the buy-and-hold strategy and a value below 1 implies that the buy-and-hold strategy outperforms our strategy.}\label{fig8}
\end{figure}

\begin{figure}
    \centering
    \rotatebox{0}{\includegraphics[width=120mm]{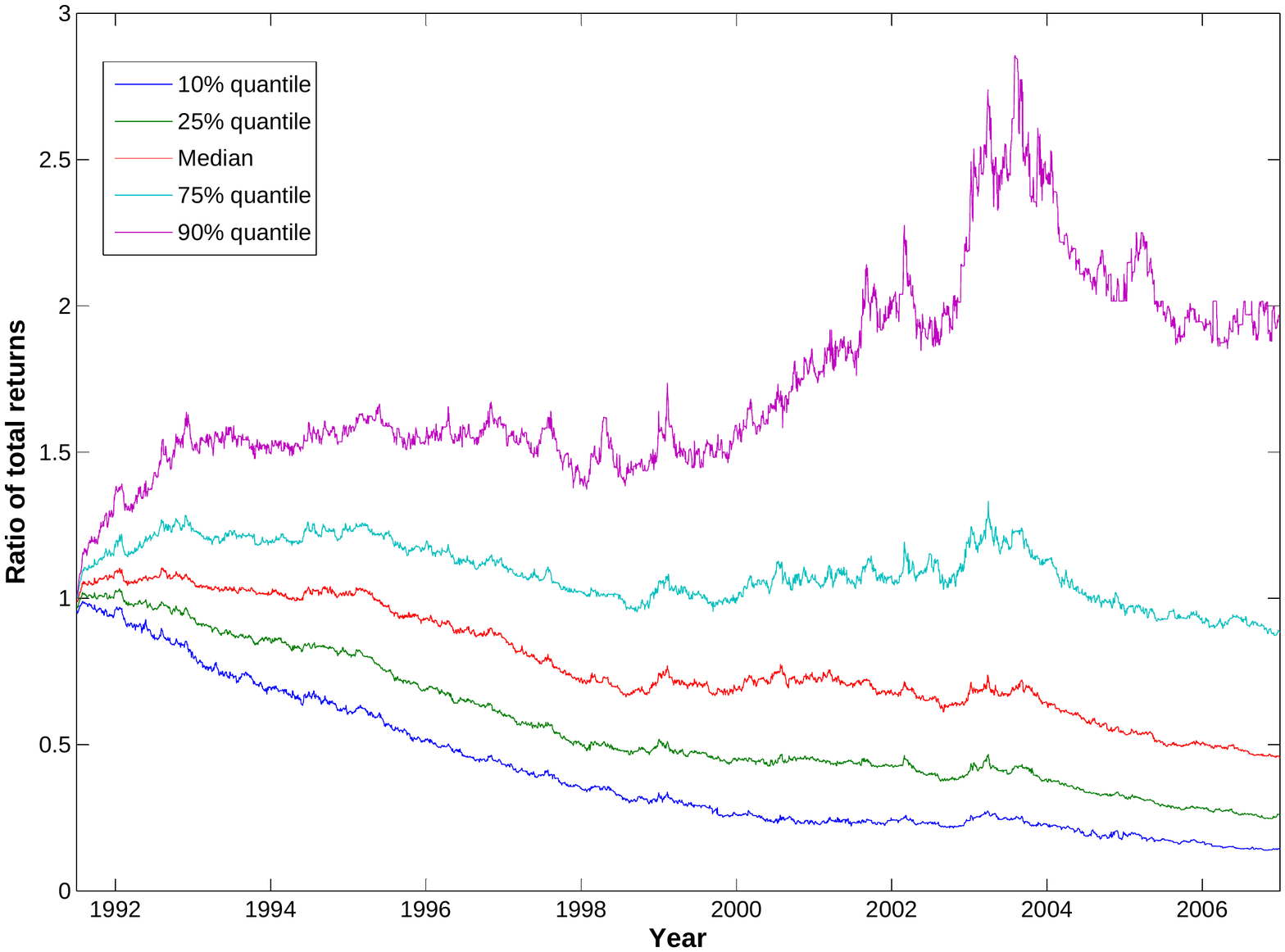}}
    \caption{Ratio of the total returns generated from our trading strategy, including trading costs on 0.1\% for each trade, and a buy-and-hold strategy from the start of the sample period. The figure shows the 10\%, 25\%, 75\%, and 90\% quantiles as well as the median of this ratio for the sample of 375 stocks considered. A value above 1 implies that our strategy outperforms the buy-and-hold strategy and a value below 1 implies that the buy-and-hold strategy outperforms our strategy.}\label{fig9}
\end{figure}

In order to assess the trading performance when using our predictions, we assume that we hold a single unit of the stock whenever the prediction states that the stock should increase and do not hold a position of the stock whenever the prediction states that the stock should decrease. Using this simple trading rule, we can now evaluate the performance of each stock. By comparing the performance of the stocks using our trading strategy with a simple buy and hold strategy, we see from figure \ref{fig8} that our strategy outperforms the buy and hold strategy for the majority of stocks, although by far not for all. Once we introduce trading costs of 0.1\% for each transaction into our model, however, our strategy produces a return which is in the majority of cases below that of the buy and hold strategy, as shown in figure \ref{fig9}.

\begin{figure}
    \centering
    \rotatebox{0}{\includegraphics[width=120mm]{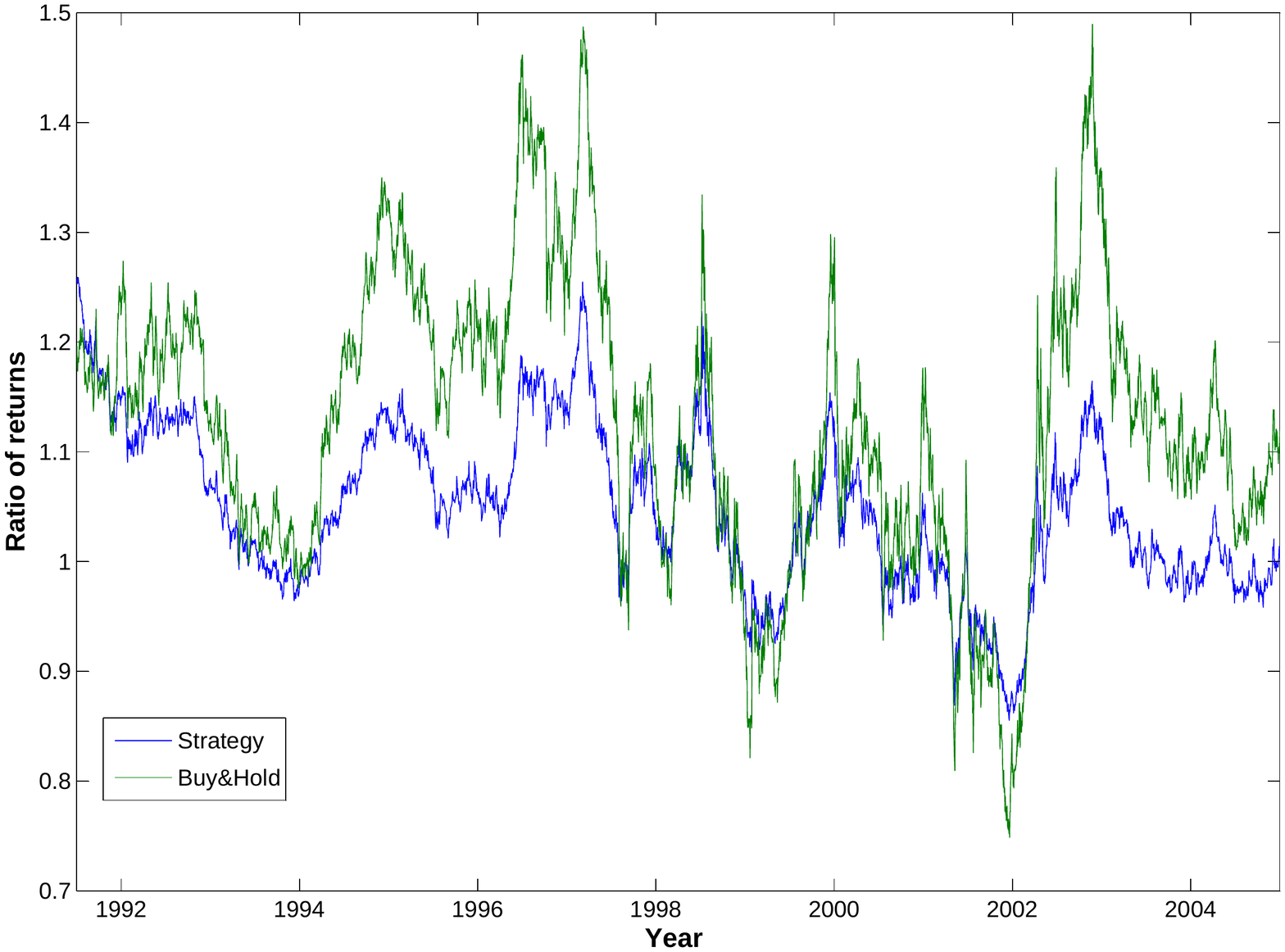}}
    \caption{Returns generated from our trading strategy, including trading costs on 0.1\% for each trade, and a buy-and-hold strategy during a 250-trading day rolling window. The figure shows the returns of an equally weighted portfolio of all 375 stocks in our sample.}\label{fig10}
\end{figure}

\begin{figure}
    \centering
    \rotatebox{0}{\includegraphics[width=120mm]{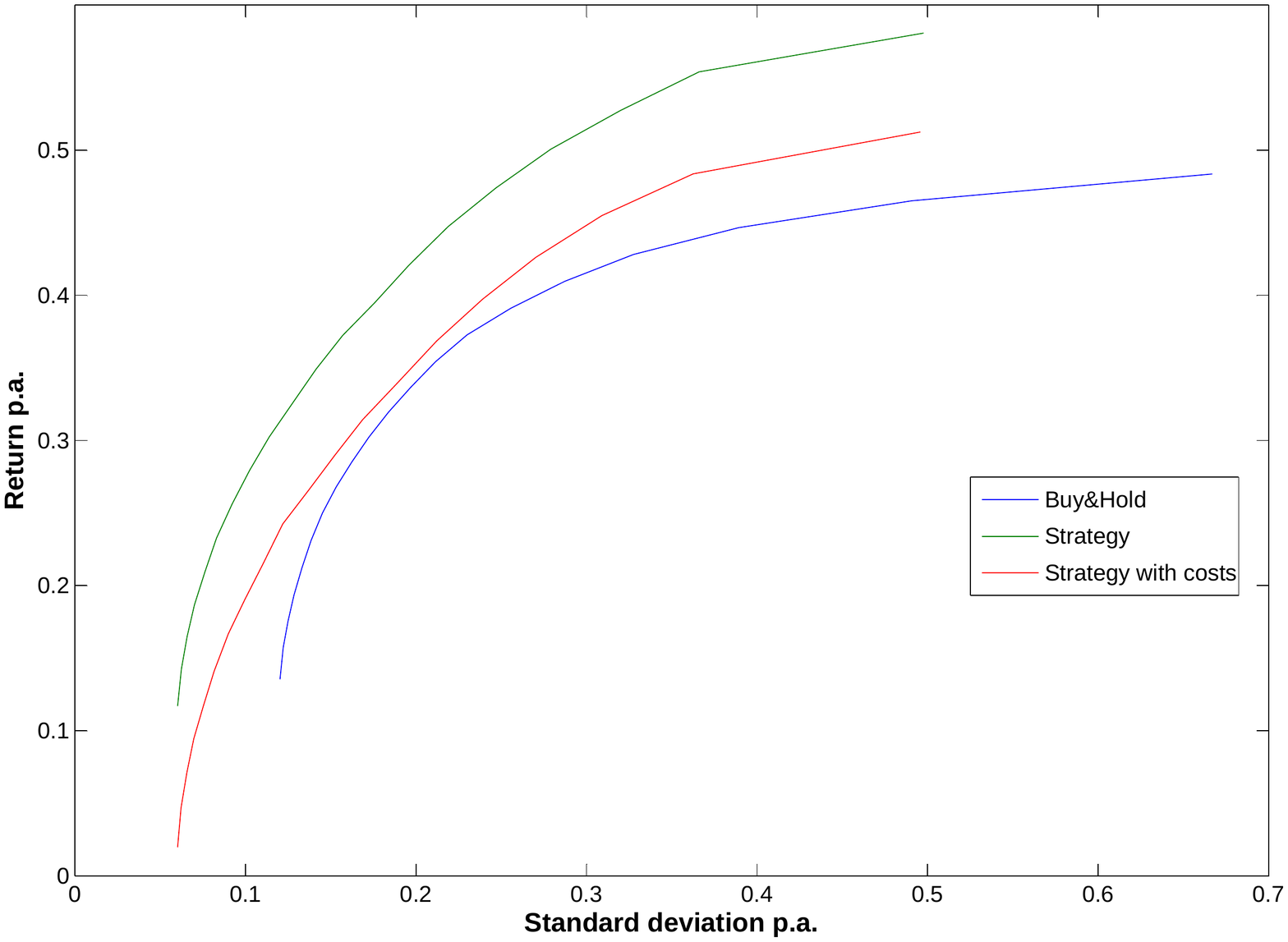}}
    \caption{Efficient frontiers generated from the 375 stocks in our sample using daily data for the entire sample period. The figure shows the median efficient frontier based on 100 sets of portfolios based on 50 randomly selected stocks.}\label{fig11}
\end{figure}

When evaluating a 250-day rolling window, i.e. comparing the performance of our strategy and a buy-and-hold strategy during one year, we see clearly from figure \ref{fig10} for the case of an equally weighted portfolio that while our strategy underperforms, it is also less volatile. Hence, a portfolio of stocks using our trading strategy might still exhibit a superior performance. In figure \ref{fig11} we compare the efficient frontiers of portfolios of 50 randomly selected stocks\footnote{We did restrict our portfolios to 50 stocks firstly as portfolios of this size are more realistic than portfolios of all 375 stocks and when investigating monthly returns we are still able to perform adequate estimations of the covariance matrices.} from our sample and we clearly see that despite the trading costs our strategy produces superior portfolios, thus showing that our trading strategy produces superior results.

\begin{figure}
    \centering
    \rotatebox{0}{\includegraphics[width=120mm]{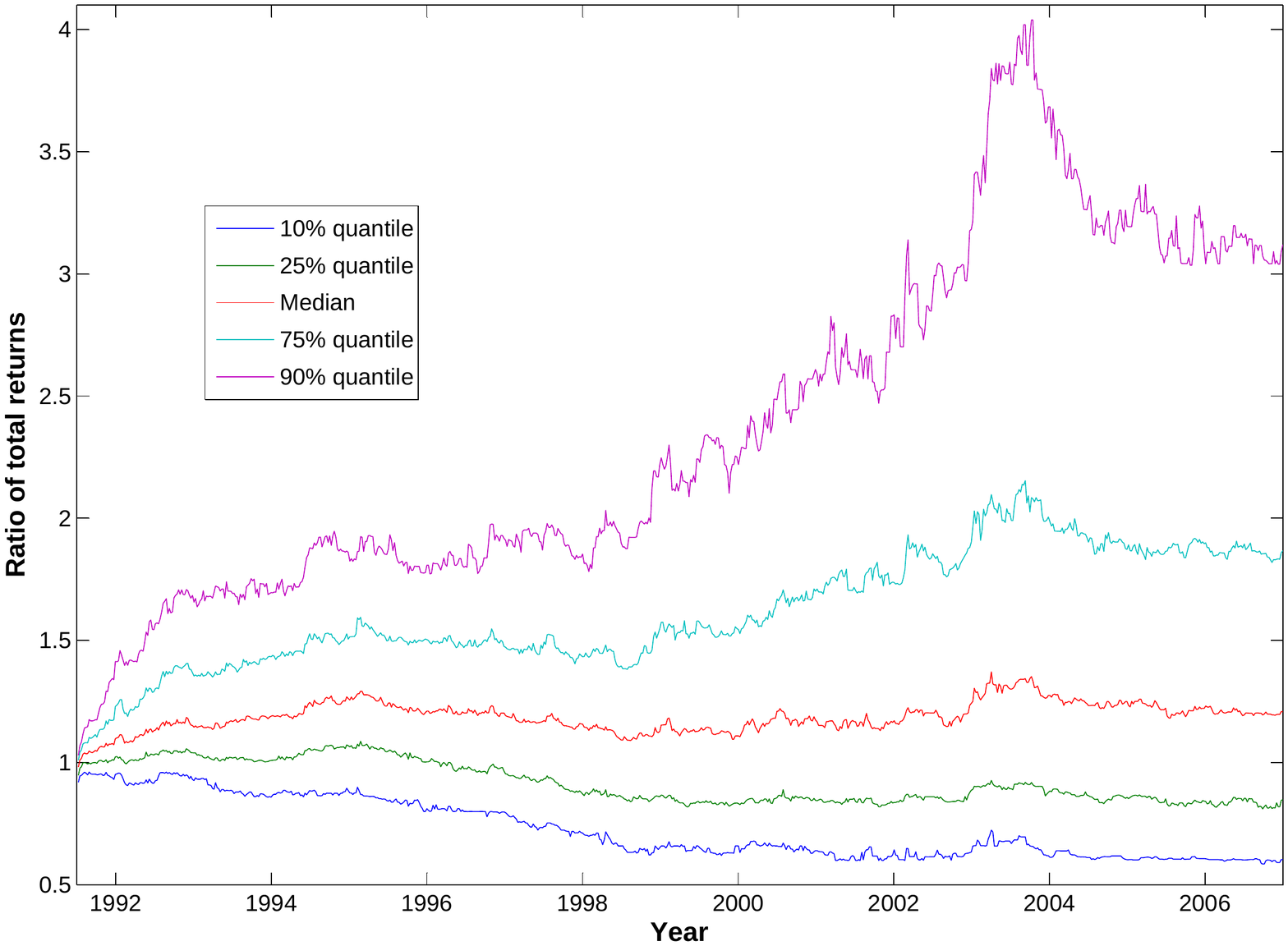}}
    \rotatebox{0}{\includegraphics[width=120mm]{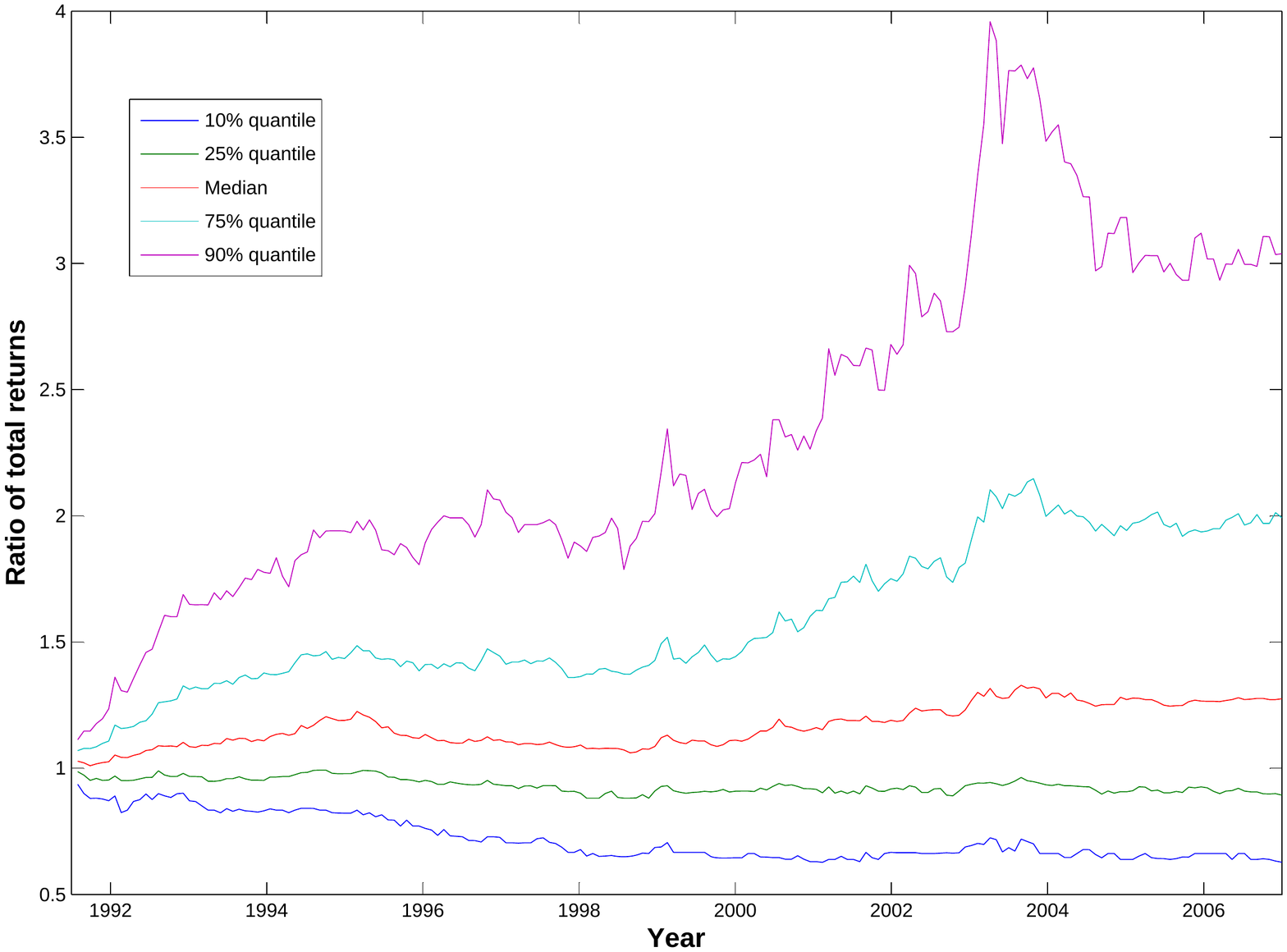}}
    \caption{Ratio of the total returns generated from our trading strategy, including trading costs on 0.1\% for each trade, and a buy-and-hold strategy from the start of the sample period using weekly (top figure) and monthly data (bottom figure). The figure shows the 10\%, 25\%, 75\%, and 90\% quantiles as well as the median of this ratio for the sample of 375 stocks considered. A value above 1 implies that our strategy outperforms the buy-and-hold strategy and a value below 1 implies that the buy-and-hold strategy outperforms our strategy.}\label{fig12}
\end{figure}

\begin{figure}
    \centering
    \rotatebox{0}{\includegraphics[width=120mm]{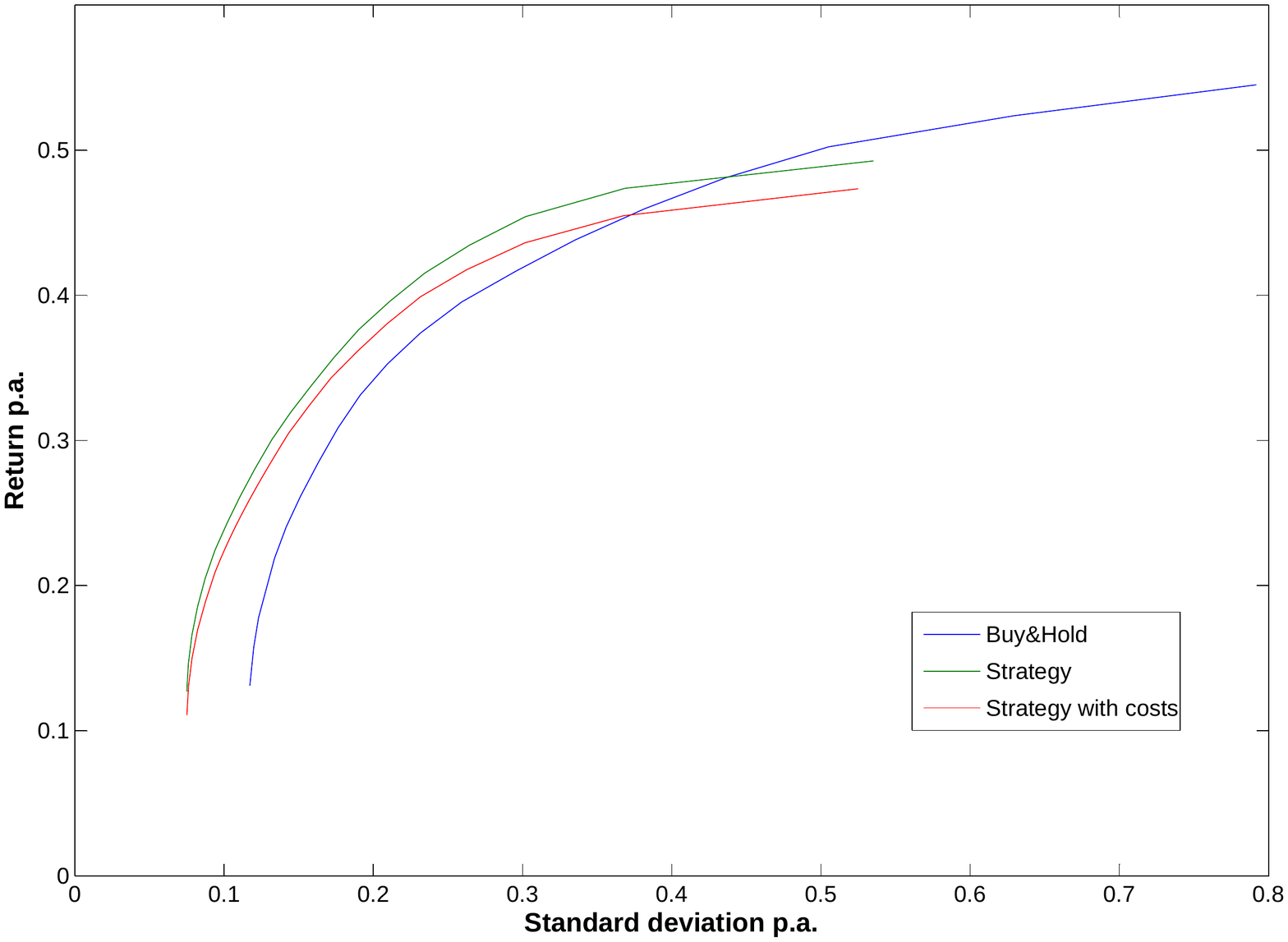}}
    \rotatebox{0}{\includegraphics[width=120mm]{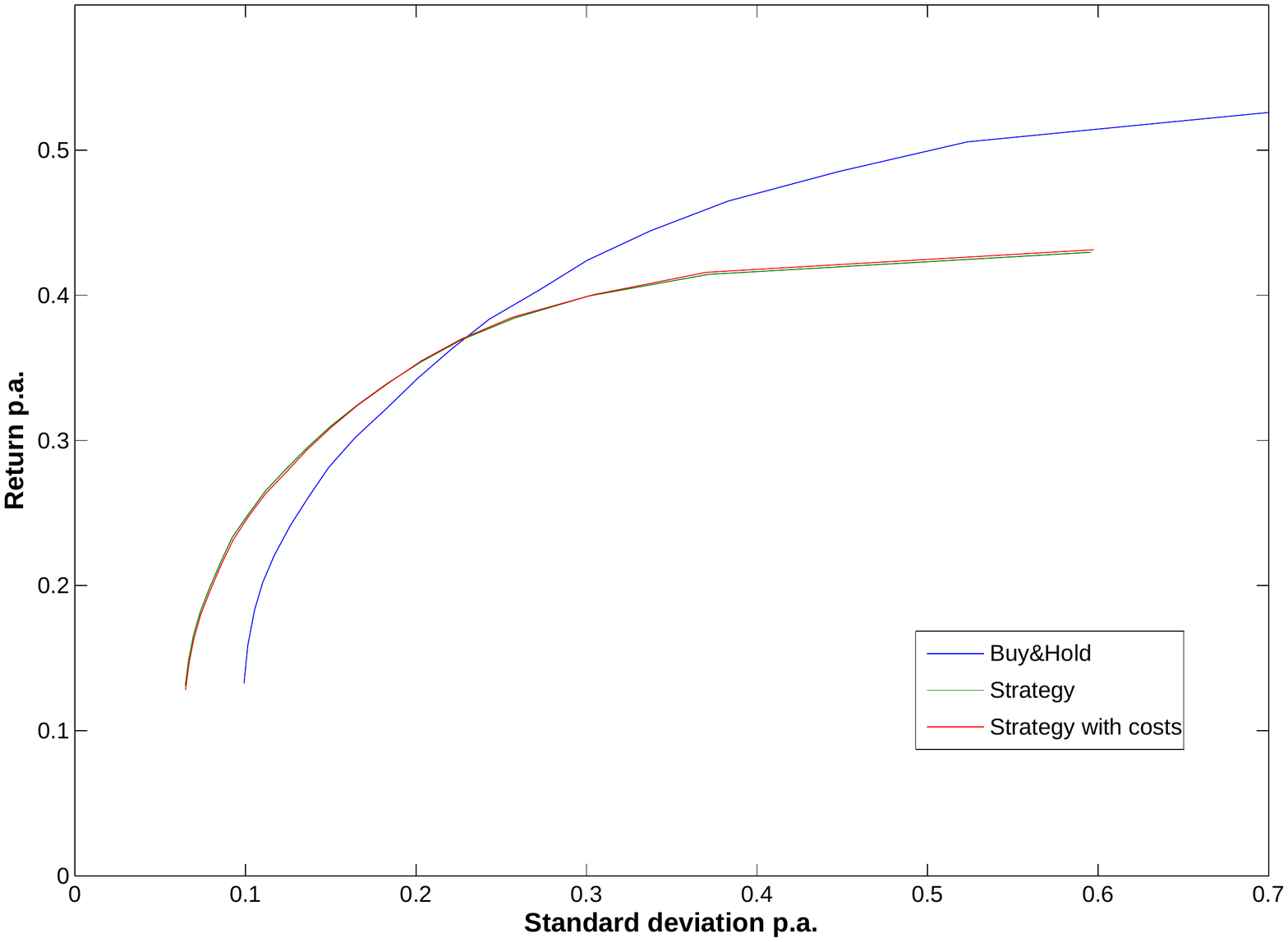}}
    \caption{Efficient frontiers generated from the 375 stocks in our sample using weekly (top figure) and monthly (bottom figure) data for the entire sample period. The figure shows the median efficient frontier based on 100 sets of portfolios based on 50 randomly selected stocks.}\label{fig13}
\end{figure}

The performance of our trading strategy is significantly reduced by the trading costs due to frequent trading. In order to reduce the impact of these trading costs we have also evaluated the trading strategy when using 5-day and 20-day returns, thus allowing to trade at most weekly or monthly, respectively. While the performance with those longer time horizons is naturally reduced when neglecting trading costs, we see from figure \ref{fig12} that the performance after trading costs is improving with the longer time horizons for an equally weighted portfolio of the stocks in our sample. Similarly we see in figure \ref{fig13} that the portfolios generated by the trading strategy are superior to those from a buy-and-hold strategy.\footnote{Please note that while the portfolios generated by the trading strategy become inferior for high returns and standard deviations, these areas become irrelevant for the optimal portfolio choice as with a risk-free rate of around 5\% p.a. the optimal risky portfolio will be well below the crossover of the efficient frontiers.}

\section{Conclusions}

We have proposed a model to predict the movements of financial assets based on ideas of the minority game. Using only past data of a given memory length we evaluated the performance of a full set of possible trading strategies across a range of memory lengths. Using the best performing strategy of the best performing memory length at the time of decision-making as the basis of a prediction for the movement of an asset in the next time period, we found that for a data set of highly liquid stocks from the S\&P500 our model was correct in 51.5-52\% of cases in any year. When investigating the trading performance we found that for the majority of stocks the buy-and-hold strategy was outperformed, however when introducing very modest trading costs this picture reversed. Nevertheless we found that the lower risk from following the trading strategy resulted in superior portfolios of stocks, thus following the trading strategy generates benefits to an investor.

The empirical evaluation of the model provides first evidence for its good performance. However, more in-depth analysis is required, e.g. by using data of shorter time scales such as intra-day data. We would also seek to investigate a wider range of assets and determine for which assets and market conditions our model is particularly suited as well as investigate whether the superior properties of the portfolios remain stable over time. Finally, we might want to consider a revised model in which past performance is discounted such that more recent outcomes receive a higher weight. We hope to address some of these aspects in future research.

\nocite{*}

\bibliographystyle{unsrt}
\bibliography{Minority}

\begin{acknowledgement}
I would like to thank Julian Williams for giving me access to his cleaned version of the S\&P500 data as well as useful discussions at an early stage of this paper.
\end{acknowledgement}

%\newpage

\end{document}